\documentclass[traditabstract]{aa}

\usepackage[english]{babel}
\usepackage{graphicx}
\usepackage{txfonts}
\usepackage{wasysym}
\newcommand{\gtap}{\mathrel{\hbox{\rlap{\lower.55ex \hbox {$\sim$}}
                   \kern-.3em \raise.4ex \hbox{$>$}}}}
\newcommand{\ltap}{\mathrel{\hbox{\rlap{\lower.55ex \hbox {$\sim$}}
                   \kern-.3em \raise.4ex \hbox{$<$}}}}
\newcommand{\izzit}{\mathrel{\hbox{\rlap{\lower.05ex \hbox {$=$}}
                   \kern-.23em \raise1.10ex \hbox{\footnotesize ?}}}}
\newcommand{\psrj}{PSR\,J0218+4232}

\begin{document}
\selectlanguage{english}
  \title{The distance and luminosity probability distributions derived from
    parallax and flux with their measurement errors}

  \subtitle{with application to the millisecond pulsar \psrj}

  \author{Andrei Igoshev \and  Frank Verbunt \and Eric Cator}

  \institute{Institute for Mathematics, Astrophysics and Particle Physics, Radboud University Nijmegen, PO Box 9010,
    6500 GL Nijmegen, The Netherlands; \email{A.Igoshev@astro.ru.nl, F.Verbunt@astro.ru.nl,  
      E.Cator@science.ru.nl}
}

%\authorrunning{A.\ Igoshev et al.}
\titlerunning{Distance and luminosity probability distributions from
  measured parallax and flux}
  \date{Draft \today}

  \abstract{We use a Bayesian approach to derive the distance
    probability distribution for one object from its parallax with
    measurement uncertainty for two spatial distribution priors, viz.\
    a homogeneous spherical distribution and a galactocentric
    distribution -- applicable for radio pulsars -- observed from
    Earth. We investigate the dependence on measurement uncertainty,
    and show that a parallax measurement can underestimate or
    overestimate the actual distance, depending on the spatial
    distribution prior.  We derive the probability distributions for
    distance and luminosity combined, and for each separately, when a
    flux with measurement error for the object is also available, and
    demonstrate the necessity of and dependence on the luminosity
    function prior. We apply this to estimate the distance and the
    radio and gamma-ray luminosities of \psrj. The use of realistic 
    priors improves the quality of the estimates for distance and luminosity,
     compared to those based on measurement only. Use of a wrong 
    prior, for example a homogeneous spatial distribution without upper bound, may lead to
    very wrong results.}


 \maketitle

\section{Introduction}

Distance determinations are fundamental in astronomy.
The study of spatial distributions and source number densities
is the most direct application. Together with proper motion
measurements, distances form the basis of velocity measurements
and kinematic studies. Combined with flux measurements
they provide luminosities. 

A standard method of distance determination is the measurement of the
trigonometric parallax. The conversion of the measured parallax into
the most probable actual parallax is not straightforward, as is
evident from the excellent historical survey given by Sandage \&\ 
Saha~(2002)\nocite{sandage2002}. Most of the papers discussed in that survey use parallax and
apparent magnitude measurements to derive absolute magnitude
distributions, or statistical corrections between apparent and
absolute magnitudes. 
In a much cited paper, Lutz \&\ Kelker (1973)\nocite{lutz1973} derive the probability
distribution of the real parallax as a function of the measured
parallax and its measurement error.
Since that paper, there has been some debate as to whether or not
their equation is applicable when only one object is observed
(as reviewed by Sandage \&\ Saha 2002).

In a study of radio pulsars, Faucher-Gigu\`ere \&\ Kaspi
(2006)\nocite{faucher2006} give a probability distribution of actual
distances as a function of the measured parallax, reproduced as
Eq.\ref{e:wrong} below.  In an important paper Verbiest et al.\
(2012)\nocite{verbiest2012} develop a Bayesian method to combine
various distance-related measurements and their uncertainties to find
the probability distribution of distances, and show the importance of
the choice of priors.  Verbiest \&\ Lorimer
(2014)\nocite{verbiest2014} apply this method in a study of the
gamma-ray luminosity of the millisecond pulsar \psrj.  Alas, they make
the same error as Faucher-Gigu\`ere \&\ Kaspi (2006) in deriving the
probability distribution of actual distances as a function of the
measured parallax, and make a similar error in the equation for the
probability distribution of luminosity as a function of measured flux
and parallax.

%In the present paper, we derive the correct equations, and show that
%they must be applied also when a single system is studied.  Indeed, a
%probability density for the distance of a single object can only be
%determined when the spatial distribution of the class of objects is
%known (or assumed).  We note that the nominal distance derived from
%the measured parallax may both under- and over-estimate the actual
%distance, depending on the spatial distribution. In this, we repeat
%some results by Bailer-Jones (2015)\nocite{bailer_jones2015} that appeared as we were
%finalizing our paper, but we differ in that we use a spatial
%distribution appropriate for pulsars.  We also derive the combined
%probability distribution of distance and luminosity, and the
%distributions of distance and luminosity separately. We show that the
%probability distribution of the luminosity can only be derived when
%the luminosity function is known or assumed, even when the flux
%measurement is accurate.

Much of the confusion in the existing literature arises because of the
failure to dicriminate between what technically are called the
frequentist approach and the Bayesian approach, leading to the
incorrect conclusion that a measurement by itself provides a
probability density distribution centered on the measured value.  
 We
briefly explain this error in Sect.\ref{s:confusion}, where we also
discuss the related confusion on whether population priors must be
taken into account in the study of single objects. In contrast to
statements in several previous papers (e.g.\ Feast 2002, Francis 2012,
and references therein), the answer is yes if a probability density is
required. A more detailed explanation is given in Sect.3.
In that
Section we repeat some results by Bailer-Jones (2015) that appeared as
we were finalizing our paper, but we differ in that we use a spatial
distribution appropriate for pulsars.

The structure of our paper is as follows. In Sect.\ref{s:ingredients}
we describe the spatial distributions and the luminosity
distributions that we use, and explain our notation.
In Section\,\ref{s:correct} we describe in some detail the derivation
of the correct conversion of measured parallax to probability
distribution of actual distances, for the case of a known (or assumed)
distribution in space. We consider a homogeneous distribution,
and a galactocentric distribution observed from Earth. 
The latter is applied to the case of \psrj. 
In Sect.\ref{s:parflux}, we consider objects for which both parallax and flux are
measured to determine the probability distributions for distance
and luminosity, and illustrate our results for \psrj.
The gamma-ray luminosity of \psrj\ is discussed 
in Sect.\ref{s:gamma}.
Finally, in Section\,\ref{s:discussion} we briefly discuss the assumptions that
we have made, and the expected consequences of relaxing these.

\subsection{Confidence intervals and probability densities}
\label{s:confusion}

Consider an object whose
parallax is measured with accuracy $\sigma$, i.e.\ the measured value
$\varpi'$ is a draw from a gaussian centered on its real parallax
$\varpi_1$ with standard deviation $\sigma$.  The probability that
a draw leads to a measured value $\varpi'$ such that
$|\varpi'-\varpi_1|<\sigma$ is then (roughly) 68\%, which corresponds to
a 68\%\ probability that the real value $\varpi_1$ is in the range given
by $\varpi'-\sigma<\varpi_1<\varpi'+\sigma$. Similarly, if the real
parallax is $\varpi_2$ there is a 68\%\ probability that the real
value $\varpi_2$ is in the range given by
$\varpi'-\sigma<\varpi_2<\varpi'+\sigma$, and so for
every real distance $\varpi_i$. Thus, no matter what the real
distance is, we can state that there is a 68\%\ probability that it is
in the range bounded by $\varpi'-\sigma$ and $\varpi'+\sigma$. 
Analogously, for each frequency of occurrence, e.g.\ expressed
in percentage $x$\%, on may derive the corresponding range:
between $\varpi'-n_x\sigma$ and $\varpi'+n_x\sigma$,
where $n_x=1.645$ for 90\%, $n_x=2$ for 95.5\%, etc.
Hence the name frequentist approach. 
The measured value $\varpi'$ does {\em not}, however, provide the
probability distribution within these ranges.

To obtain such a probability distribution one must compute the
relative contribution that each possible real parallax $\varpi_i$
makes to the probability of measuring $\varpi'$, i.e.\ follow the
Bayesian approach.  As an illustration, consider a population of 10
sources, 9 of which have $\varpi=5$\,mas, and 1 has
$\varpi=3$\,mas. We select one source from this population for a
parallax measurement with accuracy 1\,mas, and measure
$\varpi'=4$\,mas. The real parallax answers to the 68\%\ probability
of lying within 1\,mas of the measured value. A real parallax of
5\,mas has a probability of 90\%, a real parallax of 3\,mas of 10\%,
and other parallaxes have probability zero. The probability
distribution of the real distance $\varpi$ is {\em not} given by a
gaussian centered on the measured value $\varpi'$.  Also for the case
of a more realistic, continuous intrinsic distribution, the
probability distribution of $\varpi$ in general can {\em not} be
stated to be given by a gaussian centered on the measured value
$\varpi'$.  
Therefore, the use of realistic priors improves the quality of the
estimate for the distance, compared to that based on one measurement only.
The same is true for the estimate of the luminosity.

Finally, consider a series of measurements
${\varpi'}_i$ made from a single object, each with its own
accuracy $\sigma_i$. Each measurement is a draw
from a distribution centered on the actual distance of the
object. The best estimate of $\varpi'$, and its accuracy
$\sigma$ can be determined by averaging these measurements
with appropriate weighting of the individual measurements,
without reference to the population priors. The resulting
values $\varpi'$ and $\sigma$ are the best estimate of
the parallax {\em measurement} and its error. They may be used in a
frequentist approach to determine a confidence interval.
To determine a probability density, they must be combined
with a population prior.

\section{Ingredients and notation\label{s:ingredients}}

The analysis in this paper is based on measurements of parallax
and flux, combined with an intrinsic spatial distribution, which
is assumed to be known, and an intrinsic luminosity distribution,
also assumed known. The measurement errors lead to
probability distributions for measured values that we denote
with $g_D$ and $g_S$ for parallax and flux, respectively.
The intrinsic spatial and luminosity distributions are
denoted with $f_D$ and $f_L$, respectively.
To illustrate the general methods, we discuss two 
spatial distributions and two luminosity distributions. 

\subsection{Measurements}

A  parallax measurement is subject to measurement error
$\sigma$. The measurement error distribution $g_D(\varpi'|D)$ gives the
probability of measuring a parallax $\varpi'$ when the actual distance
is $D$. $g_D(\varpi'|D)$ may follow a gaussian distribution
(Eq.\ref{e:gauss}), but in general, it may also have a different,
non-gaussian form.

We will assume that the distance $D$ is given in
kiloparsecs, and the parallax $\varpi$ and measurement error $\sigma$
in milliarcsec, hence $\varpi=1/D$, and we will assume that the
parallax measurement errors follow a gaussian distribution, centered
on zero and with width $\sigma$, i.e.\ that the probability of
measuring a parallax $\varpi'$ for an actual parallax $\varpi$ is
given by a gaussian:
\begin{equation}
g_D(\varpi'|\varpi)\Delta\varpi' = {1\over\sqrt{2\pi}\sigma}
\exp\left[-\,{(\varpi-\varpi')^2\over2\sigma^2}\right] \Delta\varpi'
\end{equation}
In this equation $\varpi$ is fixed, so with $\varpi = 1/D$ we rewrite it as
\begin{equation}
g_D(\varpi'|D)\Delta\varpi' = {1\over\sqrt{2\pi}\sigma}
\exp\left[-\,{(1/D-\varpi')^2\over2\sigma^2}\right] \Delta\varpi'
\label{e:gauss}\end{equation}
where $g_D(\varpi'|D)$ is normalized over the range
$-\infty<\varpi'<\infty$. (Note that, whereas the real parallax
is by definition positive, the measured value may be negative.)
Our results for spatially homogeneous distributions
will be identical for $D$ in parsecs with
$\varpi$ and $\sigma$ in arcsecs.

We furthermore assume that the probability of a measured flux $S'$
for an actual flux $S$ is given by
\begin{equation}
g_S(S'|S)\Delta S' = {1\over\sqrt{2\pi}\sigma_S}
\exp\left[-{(S-S')^2\over2{\sigma_S}^2}\right]\,\Delta S'
\label{e:gauss-s}\end{equation}
A flux $S$ for a source at distance $D$ corresponds to a luminosity
$L=L_oD^2S$. We introduce the factor $L_o$ to
discriminate isotropically emitting sources, for which $L_o=4\pi$,
and pulsars, for which traditionally the luminosity is defined
with $L_o=1$. It may also be used to indicate the effect
of interstellar absorption, in which case $L_o$ itself depends
on $D$.

\subsection{Spatial distribution}

To avoid unnecessary duplication, we subsume the two spatial
distributions that we discuss in one equation:
\begin{equation}
f_D(D)\Delta D \propto D^2\mathcal{F}(D)\Delta D
\label{e:spatial}\end{equation}
For a homogeneous distribution in space, $f_D(D)\propto D^2$, and
$f_D(D)$ cannot be normalized.  In
realistic applications, however, the spatial distribution is
always bounded: for stars by the finite extent of the galaxy. For 
illustrative purpose, we consider the (in general non-realistic) case
where the distribution is homogeneous up to a maximum distance
$D_\mathrm{max}$, and zero beyond it; and write
\begin{equation}
\mathcal{F}(D) =  \left\{ \begin{array}{cc}
 1 & \mathrm{for}\,D<D_\mathrm{max} \\
\\
0 & \mathrm{for} \, D>D_\mathrm{max}
\end{array}\right. \label{e:homodist}\end{equation}

Verbiest et al.\,(2012) consider the observations made from Earth on a
galactocentric distribution, which results in a heliocentric
distribution given in our notation by (cf.\ Eq.\,21 of Verbiest et al.\,2012):
\begin{equation}
\mathcal{F}(D)= R^{1.9}\exp\left[
-{|z(D,b)|\over h}-{R(D,l,b)\over H}\right]
\label{e:galacent}\end{equation}
Here a cylindrical galactocentric coordinate system is adopted with
$R$ and $R_o$ the distance of the pulsar and of Earth to the galactic
center, projected onto the galactic plane, and $z$ the distance of the
pulsar to that plane. $h$ and $H$ are the vertical and radial scaling
parameters. With $D$ the distance of the object to Earth, and $l,b$ its
galactic coordinates, we have
\begin{equation}
z = D\sin b;\,\mathrm{and}\, R=\sqrt{{R_o}^2+(D\cos b)^2-2D\cos b\,R_o\cos l}
\end{equation}
The last equation shows that $z$ and $R$ are functions of $D$, $l$ and $b$.
and thus $\mathcal{F}(D)$ and through it $f_D(D)$ are functions of $l$ and $b$.

\subsection{Luminosity functions}

The luminosity function $f_L(L)$ gives the relative numbers of sources
as a function of luminosity $L$, in the range between minimum 
luminosity $L_\mathrm{min}$ and maximum luminosity $L_\mathrm{max}$.
The luminosity function 
$f_L(L)$, and also $L_\mathrm{max}$ and  $L_\mathrm{min}$,  may 
depend on $D$. For example, pulsars at large distance from the
galactic plane tend to be  older, and probably have a luminosity function 
different from that of young pulsars near the galactic plane. 
However, for the purpose of this paper, we assume a universal luminosity
function, i.e.\ $f_L(L)$, $L_\mathrm{min}$ and $L_\mathrm{max}$
do not depend on $D$.

As a first example we discuss a power-law distribution
for the luminosity function:
\begin{equation}
f_L(L)\Delta L \propto \left\{ \begin{array}{cc}
 L^\alpha \Delta L
  & \mathrm{for}\,L_\mathrm{min}<L<L_\mathrm{max} \\
\\ 0 & \mathrm{for} \, L>L_\mathrm{max}\,\mathrm{or}\,
L<L_\mathrm{min}.\end{array}\right. 
\label{e:powerlaw}\end{equation}
We will consider three values for $\alpha$, viz. $\alpha=(0, -1, -2)$.

We also consider a luminosity function in the form derived for
normal pulsars by Faucher-Gigu\`ere \&\ Kaspi (2006):
\begin{equation}
f(x)\Delta x \propto
\exp\left[-\,{(x-\mu_x)^2\over2{\sigma_x}^2}\right]\Delta x
\quad\mathrm{where}\quad x\equiv\log L
\end{equation}
which we rewrite as
\begin{equation}
f_L(L)\Delta L = f(x){dx\over dL}dL \propto
\exp\left[-\,{(\log L-\mu_x)^2\over2{\sigma_x}^2}\right]{1\over L}dL
\label{e:lognormal}\end{equation}
where $\mu_x=-1.1$ and $\sigma_L=0.9$ (both numbers
referring to the log of the luminosity in mJy\,kpc$^2$). 
We follow Verbiest et al.\ (2014) in applying this same distribution
to millisecond pulsars.

\subsection{Notation for probabilities}

We denote joint probabilites with capital $P$, in particular
the joint probability of measured parallax $\varpi'$ and actual
distance $D$ is written $P(\varpi',D)$, and the joint probability for
these quantities plus measured flux $S'$ and luminosity $L$
as $P(\varpi',D,S',L)$. These joint probabilities may be turned
into conditional probabilities with Bayes' theorem. This leads to
normalization constants which we denote as follows.
If the joint probability is
\begin{equation}
P \propto F(\varpi',D,S',L)
\end{equation}
with $F$ a function of the variables indicated, then
the conditional probability
\begin{eqnarray}
p &=& C_x(a,b) F(\varpi',D,S',L); \quad x\in\,\{\varpi',D,S',L\}
\nonumber\\
& &\mathrm{with}\quad {C_x(a,b)}^{-1} \equiv \int_a^b F(\varpi',D,S',L)dx
\label{e:bayes}\end{eqnarray}
Our notation for conditional probabilities is
such that
\begin{equation}
p_x(x|a,b\ldots)
\label{e:conditional}\end{equation}
gives the (normalized) probability of $x$ for given
(e.g.\ measured)  values for $a, b,\ldots$.

We will use 95\%\ credibility intervals on the posterior
 probability density.
This credibility
interval is computed from the one-dimensional posterior
probability density $p_x(x)$, where $x$ is the distance or the luminosity,
as the shortest interval containing 95\%\ of the total probability:
\begin{equation}
\int_{x_l}^{x_u}p_x(x)dx=0.95\int_0^\infty p_x(x)dx\qquad\mathrm{with}\hspace*{0.25cm}
p_x(x_l)=p_x(x_u)
\label{e:credibility}\end{equation}

This equation holds when $x_u<x_\mathrm{max}$; when
  $x_u=x_\mathrm{max}$, the condition $p_x(x_l)=p_x(x_u)$
  is dropped.

\begin{table}
\caption{Parameters of \psrj\ used in this paper\label{t:psrj}}
\begin{tabular}{l@{\hspace{0.05cm}}c@{\hspace{0.1cm}}c@{\hspace{0.1cm}}l}
\hline
\multicolumn{3}{c}{specific for \psrj} & reference \\
coordinates & $l,b$ & $139\fdg51,-17\fdg53$ \\
\multicolumn{3}{l}{period, -derivative\quad $P$,$\dot P$ \, 2.323\,ms, $7.739\times10^{-20}$}  & (1) \\
parallax & $\varpi',\sigma$ & $0.16\pm0.09$\,mas & (2) \\
flux\,1400\,Mhz  & $S',\sigma_S$ & $0.9\pm0.2$\,mJy & (3) \\
flux\,0.1-100\,GeV & $S_\gamma$ & \hspace*{-0.5cm} 
$4.56\times10^{-11}\mathrm{erg\,s}^{-1}\mathrm{cm}^{-2}$ & (4)\\
\hline
 \multicolumn{3}{c}{generic for millisecond pulsars} & reference \\
in Eq.\ref{e:galacent} & $R_o,H,h$ & 8.5\,kpc,0.2$R_o$,500\,pc & (5) \\
Eq.\ref{e:lognormal} $L_{1400}$ & $\mu_x,\sigma_x$ & $-1.1,0.9^a$ & (6)\\
{ Eq.\ref{e:lognormal} $L_\gamma$} & $\mu_x,\sigma_x$ & $32.7,1.4^b$ & (7)\\
\hline
\end{tabular}
\tablefoot{ $^a$actually derived for normal pulsars; both numbers refer to the log of the luminosity
  in mJy\,kpc$^2$; $^b$ both numbers refer to the log of the gamma-ray
  luminosity in erg/s. 

References are: (1) -- \cite{hobbs2004}, (2) -- \cite{du2014}, (3) -- \cite{kramer1998}
(4) -- \cite{abdo2013}, (5) -- \cite{lorimer2006}, (6) -- \cite{faucher2006}
and \cite{hooper2015}.
}
\end{table}

\subsection{Sample millisecond pulsar}

In the Figures which illustrate probabilties involving
the galactocentric distribution Eq.\ref{e:galacent}
we will use the parameters for \psrj, as listed in
Table\,\ref{t:psrj}.

\section{Distance derived from measured parallax and
assumed distance distribution\label{s:correct}}

Due to the measurement error, different distances $D$ may lead to the
same measured parallax $\varpi'$.  With the number of objects at
distance $D$ given by $f_D(D)$, and the
probability of measuring parallax $\varpi'$ at actual distance $D$ 
by $g_D(\varpi'|D)$, the joint probability of a object
to have a distance $D$ and a measured parallax $\varpi'$ is
distributed according to
\begin{equation}
 P_D(\varpi',D) \Delta\varpi'\,\Delta D= 
 g_D(\varpi'|D)\,f_D(D)\Delta\varpi'\,\Delta D
\label{e:joint}\end{equation}
and the conditional probability that the actual distance is in a range
$\Delta D$ around $D$ when the measured parallax is 
$\varpi'$ follows with Eq.\ref{e:bayes}:
\begin{equation}
p_D(D|\varpi')\Delta D = C_D(0,\infty) P_D(\varpi',D)\Delta D
\label{e:jointnorm}\end{equation}
In principle, only the product $P_D=g_Df_D$ must be normalizable with respect
to $D$; in practice it is often useful to normalize the functions $g_D$ and $f_D$ separately
as well, with respect to $D$ and $\varpi'$, respectively.
Eqs.\ref{e:joint} and \ref{e:jointnorm} show that a probability
distribution for the distance can be derived for a measured parallax
of a single object only if a spatial distribution $f_D(D)$ of the class of objects
is known or assumed.     

For a uniform prior, i.e.\ $f_D(D)=\mathrm{constant}$ in
the range $D_\mathrm{min}<D<D_\mathrm{max}$,
Eqs.14-15 lead to the result
\begin{equation}
p_D(D|\varpi') = C_D(D_\mathrm{min},D_\mathrm{max}) g_D(\varpi'|D) 
\label{e:uniprior}\end{equation}
Thus, for a uniform prior, the probability of measuring $\varpi'$
when the real distance is $D$ is the same as the probability that the
real distance is $D$ when the measured parallax is $\varpi'$, apart
from a normalization constant.
To prevent the normalization constant from going to infinity,
the prior may have to be limited to a maximum distance.

\subsection{Finite homogeneous distribution in space}

\begin{figure}
\centerline{\includegraphics[angle=0,width=\columnwidth]{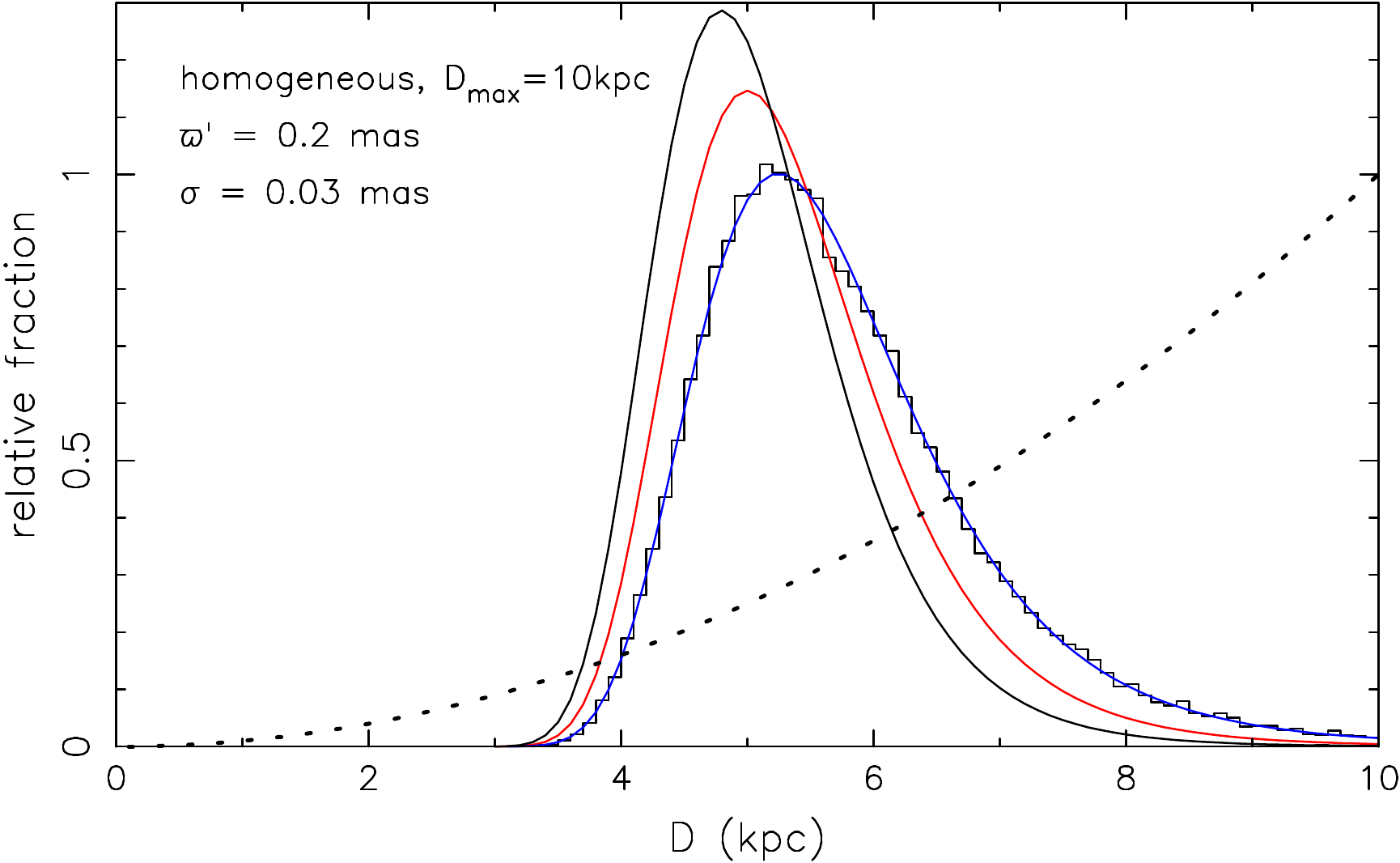}}

\caption{The probability distribution of actual distances for a measured parallax
  for  objects distributed homogeneously in a finite sphere; for
  values of $\varpi'$, $\sigma$ and $D_\mathrm{max}$ as
  indicated. The blue line represents
  Eq.\ref{e:normhomogen}. The histogram gives the results of a Monte
  Carlo simulation which retains objects with $0.198<\varpi'<0.202$.
  The black and red line represent modified versions of
  Eq.\ref{e:normhomogen} according to
  Faucher-Gigu\`ere \&\ Kaspi (2006) and Verbiest et
  al.\ (2012), respectively. The intrinsic distribution given by
  Eqs.\ref{e:spatial},\ref{e:homodist} is shown as a dashed line.
 All curves are normalized to the same  area under the curve %{\bf lutzk.f90}
 \label{f:homogen}}
 \end{figure}

\begin{figure}
\centerline{\includegraphics[angle=0,width=\columnwidth]{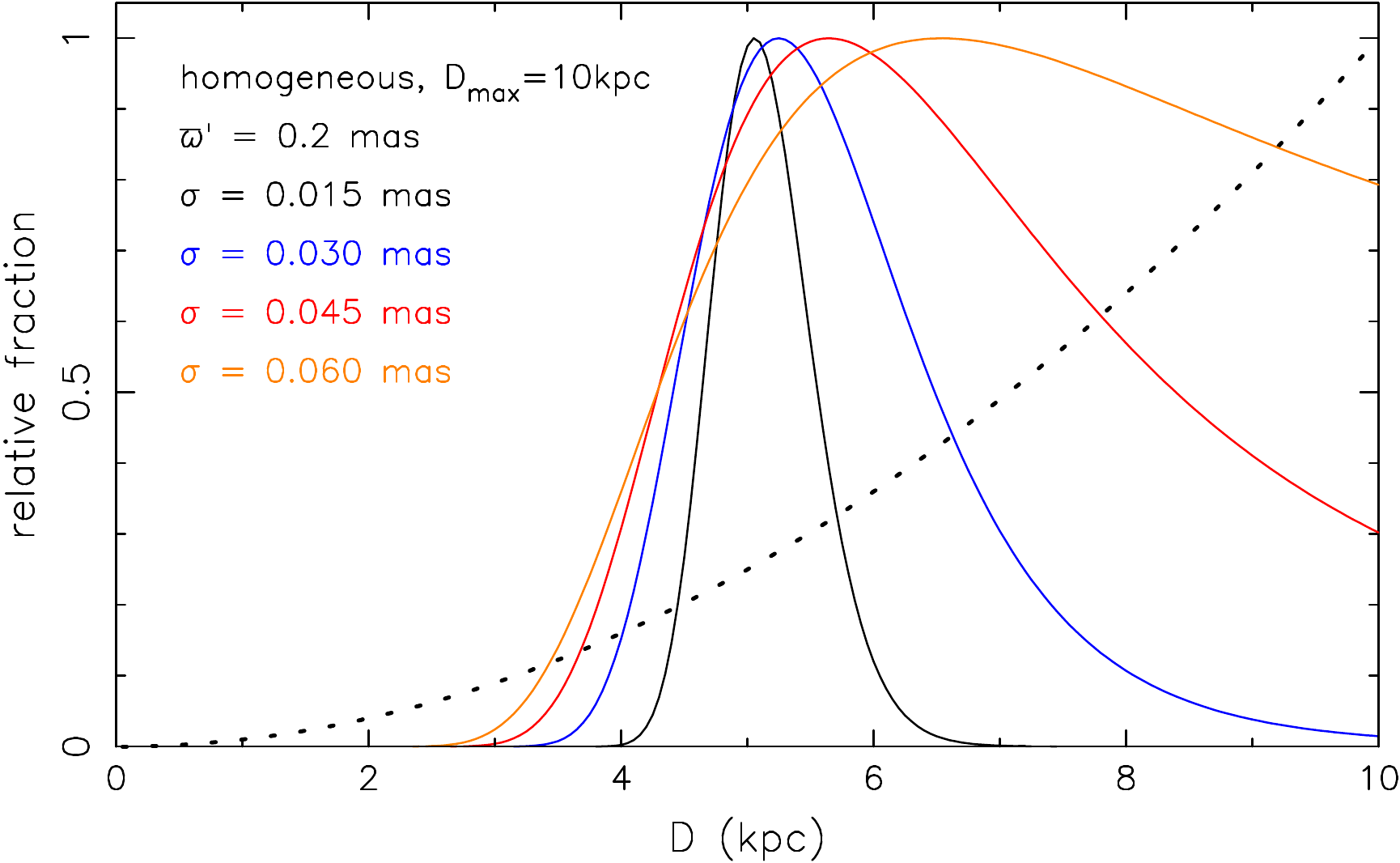}}

\caption{The probability distribution of actual distances for a measured parallax
  $\varpi'=0.2$\,mas for various measurement errors $\sigma$, for
  objects distributed homogeneously in a sphere with radius 
  $D_\mathrm{max}=10$\,kpc. The intrinsic distribution given by
  Eqs.\ref{e:spatial},\ref{e:homodist} is shown as a dashed line.
  The curves are normalized to the same maximum value %{\bf lutzkb.f90}
 \label{f:homogenb}}
 \end{figure}

Entering Eqs.\ref{e:gauss}, \ref{e:spatial}, \ref{e:homodist} into
Eq.\ref{e:jointnorm}, we obtain with Eq.\ref{e:bayes}:
\begin{equation}
p_D(D|\varpi') = \left\{ \begin{array}{cc}
C_D(0,D_\mathrm{max})
D^2\exp\left[-\,{(1/D-\varpi')^2\over2\sigma^2}\right];
\,D<D_\mathrm{max}  \\
 \\
 0;\quad  D>D_\mathrm{max} 
\end{array}\right. 
\label{e:normhomogen} 
\end{equation}
In Fig.\ref{f:homogen} we plot $p_D(D|\varpi')$ according to
Eq.\ref{e:normhomogen}, computing $C_D(0,D_\mathrm{max})$
numerically, for a measured parallax $\varpi'=0.2$\,mas, maximum
distance $D_\mathrm{max}=10$\,kpc and $\sigma=0.03$\,mas.
Fig.\ref{f:homogenb} illustrates the effect of varying measurement
accuracies. As the error decreases, the most probable distance
closes in to the nominal measured value $1/\varpi'$, but the probability
distribution of the actual distances remains asymmetric, i.e.\
non-gaussian, even for small measurement errors.

To show that our approach is in agreement with that of Lutz \&\ Kelker
(1973), we note that for a homogeneous distribution in space
$f_\varpi(\varpi)\Delta\varpi=f_D(D)\Delta D\propto D^2\Delta D$,
hence 
$f_\varpi(\varpi) \propto \varpi^{-2}d(1/\varpi)/d\varpi \propto
\varpi^{-4}$. 
This allows us to write
the joint probability of a pulsar to have measured parallax $\varpi'$
and actual parallax $\varpi$ analogous to Eq.\ref{e:joint} as
\begin{eqnarray}
P_\varpi(\varpi',\varpi)\Delta\varpi'\Delta\varpi&=&g_\varpi(\varpi'|\varpi)
\Delta\varpi'\,f_\varpi(\varpi)\Delta\varpi
\nonumber \\
&\propto& {\Delta\varpi\over\varpi^4} g(\varpi'|\varpi)\,\Delta\varpi'
\end{eqnarray}
thus confirming the $\varpi^{-4}$ dependence found by Lutz \&\ Kelker.

\subsection{Galactocentric distribution}

\begin{figure}
\centerline{\includegraphics[angle=0,width=\columnwidth]{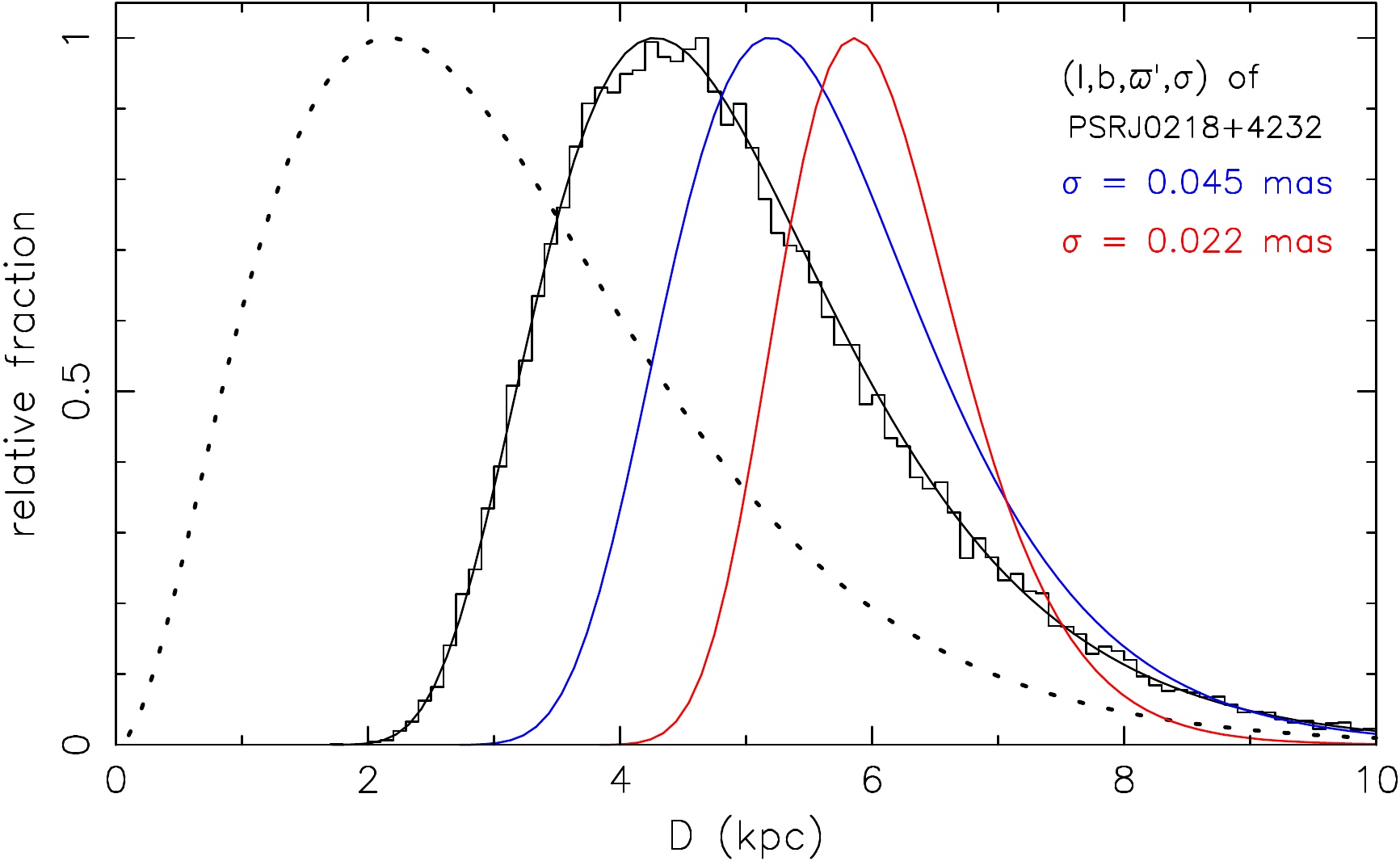}}

\caption{For an assumed galactocentric distribution of objects, the
distribution  as a function of distance to Earth is given by 
Eqs.\ref{e:spatial},\ref{e:galacent}, illustrated for the direction towards 
\psrj\ with the dotted line.
The black smooth line gives the probability distribution of actual distances in this
direction for the measured parallax of this pulsar,
according to Eq.\ref{e:jointgal}, in the approximation
$C_D(0,\infty)\simeq C_D(0,D_\mathrm{max})$, $D_\mathrm{max}=10$\,kpc.
The histogram gives the results of a Monte
Carlo simulation which retains objects with $0.14<\varpi'<0.18$.
The blue and red lines give the analytic distributions for hypothetically
smaller measurement errors but the same value for $\varpi'$.
The curves are normalized to the same maximum value  
% {\bf distlb.f90}
\label{f:galacent}}
 \end{figure}

Entering Eqs.\ref{e:gauss}, \ref{e:spatial}, \ref{e:galacent} into
Eq.\ref{e:jointnorm}, we obtain with Eq.\ref{e:bayes}:
\begin{eqnarray}
p_D(D|\varpi') &=& C_D(0,\infty) D^2\,R^{1.9}\exp\left[
-{|z(D,b)|\over h}-{R(D,l,b)\over H}\right]\nonumber \\
& & \times \exp\left[-\,{(1/D-\varpi')^2\over2\sigma^2}\right] 
\label{e:jointgal}\end{eqnarray}
Fig.\ref{f:galacent} illustrates this distribution for the parameters of \psrj.

\subsection{Earlier studies\label{s:wrong}}

Previous authors have given different expressions for
$p_D(D|\varpi')$.  To understand the difference between our
Eqs.\ref{e:normhomogen},\ref{e:jointgal} and these expressions, we
consider the measurement process expressed in Eq.\ref{e:joint}.
Consider a class of objects distributed in space according to $f_D(D)$
(Eq.\ref{e:spatial}). The measurement process starts with the
selection of one object whose parallax we wish to measure.  This
corresponds to taking a draw from the $f_D(D)$ distribution.  Then the
parallax is measured. The measurement refers to the unique distance
$D$ of the selected object, i.e.\ the selection from $g_D(\varpi'|D)$
is taken for a unique and fixed value of $D$.

We illustrate this separation between object selection and
parallax measurement with a Monte Carlo experiment, as follows.  We
choose a distance $D$ randomly from a $D^2$ distribution
(corresponding to a homogeneous distribution in a sphere) with maximum
distance 10\,kpc; for the distance $D$ a measured parallax $\varpi'$
is drawn randomly from a Gaussian distribution according to
Eq.\ref{e:gauss} with $\sigma=0.03$\,mas.  We retain the distance if
$0.198<\varpi' \mathrm{(mas)}<0.202$, and repeat the procedure until
$50\,000$ distances are retained.  The binned distribution of the
distances $D$ of the retained objects is normalized and also plotted
in Fig.\ref{f:homogen}.  It agrees with Eq.\ref{e:normhomogen}.
In analogous fashion we  perform a Monte-Carlo experiment for the 
galactocentric distribution, for parameters of the millisecond pulsar 
\psrj, and show in Fig.\ref{f:galacent} that the result
agrees with the analytic solution given by Eq.\ref{e:jointgal}.

Faucher-Gigu\`ere \&\ Kaspi (2006) write the probability
of distance $D$ for a measured parallax $\varpi'$ as
(see their Eq.\,2):
\begin{equation}
p(D|\varpi') ={C\over D^2}\, {1\over\sqrt{2\pi}\sigma}
\exp\left[-\,{(1/D-\varpi')^2\over2\sigma^2}\right] 
\label{e:wrong}\end{equation}
where $C$ (in our notation) is the normalization constant.
In doing so they make two, related, errors.
First, they interpret the right hand side of Eq.1 as giving the
probability that the real parallax is $\varpi$ when the measured value
is $\varpi'$, when in fact it gives the probability of measuring
$\varpi'$ when the real parallax is $\varpi$. As we explain in
Sect.1, this is incorrect, and arises from confusing
the frequentist and Bayesian methods. Second, by interpreting
the right hand side of Eq.1 as a probability density for $\varpi$,
they add the factor $|d\varpi/dD|=1/D^2$ in converting this to 
a probability density for $D$; and ignore the spatial density
$f(D)$. As may be seen from Eqs.2, 14 and 15 this corresponds
effectively to assuming $f(D)\propto 1/D^2$.
%two errors. First, they ignore the intrinsic
%distance distribution ($f_D(D)$ in our notation).  Second, they add an
%extra term $|d\varpi/dD|=1/D^2$ to measurement distribution
%Eq.\ref{e:gauss}. The fallacy in the latter follows from the fact
%that the measurement process relates to the unique {\em actual
%  distance}, and Eq.\ref{e:gauss} gives the distribution of {\em
%  measured} parallaxes $\varpi'$; i.e.\ involves $d\varpi'$. 
The
effect of this double error for a homogeneous spatial
distribution is to replace the $D^2$ factor in our
Eqs.\ref{e:normhomogen} with $D^{-2}$, and is
illustrated in Fig.\ref{f:homogen}.

Verbiest et al.\ (2012) and Verbiest \&\ Lorimer (2014) 
make the same errors as Faucher-Gigu\`ere and Kaspi (2006),
but correctly include $f(D)$ into the probability density
$P_D(D|\varpi')$. The net effect of this
%correct the
%first error of Faucher-Gigu\`ere \&\ Kaspi (2006), but retain the
%second. The effect of the second error 
is to remove the $D^2$ factor
in our Eqs.\ref{e:normhomogen} and~\ref{e:jointgal}, which correponds to the assumption of a uniform distance
 distribution $f_D=1$. The result is illustrated
in Fig.\ref{f:homogen}  for a homogeneous spatial distribution.

Francis (2014) argues that the distance probability distribution is a 
Gaussian centered on the real value, because it collapses to the
real value when the measurement error goes to zero.
The effect of the spatial distribution prior $D^2\mathcal{F}$ 
does diminish when the parallax measurement error becomes smaller, 
because a smaller range of $D$ leads to a smaller variation 
of the prior $D^2\mathcal{F}$. Thus, for smaller errors the distance
probability distribution narrows towards the correct distance.
However, even for small errors, the distance probability function
remains asymmetric (Figs.\ref{f:homogenb} and \ref{f:galacent}).
%Thus, the argument by Francis (2014)\nocite{francis2014} is wrong, as is his conclusion
%that the prior should not be considered in the study of a single object..
Indeed, Eq.3.2 from Francis (2014) is wrong, and confuses the 
frequentist and Bayesian approach, as does his conclusion that
the distance distribution is irrelevant for the derivation of the
probability density for the distance.

\section{Distance and luminosity from parallax and flux, with
 assumed distance and luminosity distributions\label{s:parflux}}

We now consider sources for which parallax and flux have been measured,
and the spatial distribution and luminosity function are known or assumed.
The joint probability for $D,\varpi',L,S'$ may be written
with Eqs.\ref{e:gauss},\ref{e:spatial},\ref{e:gauss-s} as
%\begin{eqnarray}
% P(D,\varpi',L,S') & \propto&
%g_D(\varpi'|D)f_D(D)g_S(S'|S[L,D])f_L(L) 
%\nonumber \\
% & = & {1\over\sqrt{2\pi}\,\sigma}
%\exp\left[-\,{(1/D-\varpi')^2\over2\sigma^2}\right]
% D^2\mathcal{F}(D)  \times \nonumber \\
% & &  {1\over\sqrt{2\pi}\,\sigma_S}
% \exp\left[-{(L/[L_oD^2]-S')^2\over2{\sigma_S}^2}\right] \, f_L(L)
%\label{e:jointall}\end{eqnarray}

$$ P(D,\varpi',L,S')  \propto
g_D(\varpi'|D)f_D(D)g_S(S'|S[L,D])f_L(L)  =$$
$$\phantom{.}\hspace{1.cm}  {1\over\sqrt{2\pi}\,\sigma}
\exp\left[-\,{(1/D-\varpi')^2\over2\sigma^2}\right]
 D^2\mathcal{F}(D)  \times $$
\begin{equation}
\phantom{.}\hspace{2.cm}  {1\over\sqrt{2\pi}\,\sigma_S}
 \exp\left[-{(L/[L_oD^2]-S')^2\over2{\sigma_S}^2}\right] \, f_L(L)
\label{e:jointall}
\end{equation}

\begin{figure}
\centerline{\includegraphics[angle=0,width=\columnwidth]{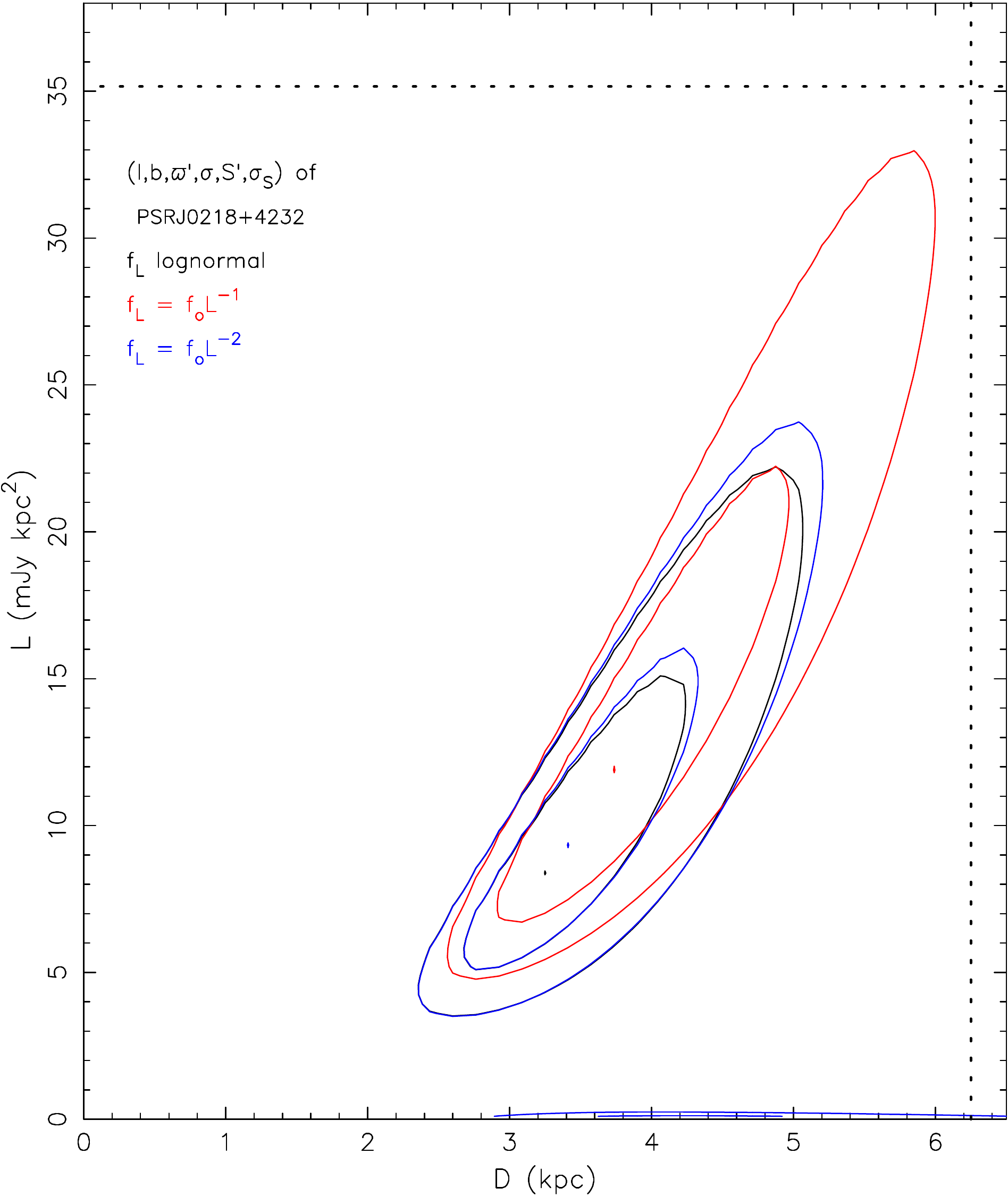}}

\caption{Contours of equal joint probability $P(D,\varpi',L,S')$, for
the direction, parallax, flux, and measurement errors of \psrj, for 
two power-law and for the lognormal luminosity functions. For each
luminosity function we show the maximum, and contours
containing 68\%\ and 95\%\ of the integrated probability.
The contours for the power-law luminosity function with index $-2$  have two branches,
one at very low luminosities, and one at higher luminosities; part of the
latter is indistinguishable from the contours for the 
the lognormal luminosity function.
The vertical and horizontal dashed lines show the nominal values for
distance $D=1/\varpi'$ and luminosity $L'=S'/{\varpi'}^2$, respectively
  \label{f:dandl}}
 \end{figure}

For fixed values of $\varpi'$, $\sigma$, $S'$ and $\sigma_S$,
and for a chosen luminosity function $f_L(L)$, this joint
probability can be computed for each combination of $D$ and $L$.
We show contours of equal probability in the $D,L$-plane in
Fig.\ref{f:dandl}, as applicable to \psrj.
The maximum probabilities lie at distances well below the 
nominal distance $D'=1/\varpi'$ and at luminosities well below the
nominal luminosity $L'=L_oS'/{\varpi'}^2$. This is due to the 
luminosity functions, that peak at values well below $L'$,
and thus favour low luminosities, hence small distances, as
far as the measurement uncertainties allow.

In Fig.\,\ref{f:dandl} we did not apply cutoffs to the power-law
luminosity functions at low or high luminosity. As may be seen
from Eq.\ref{e:jointall} such cutoffs do not change the form 
of the contours of $P(D,\varpi',L,S')$, but only the normalization,
in the range $L_\mathrm{min}<L<L_\mathrm{max}$. Outside this range
$P(D,\varpi',L,S')=0$.

\subsection{Distances}

\begin{figure}
\centerline{\includegraphics[angle=0,width=\columnwidth]{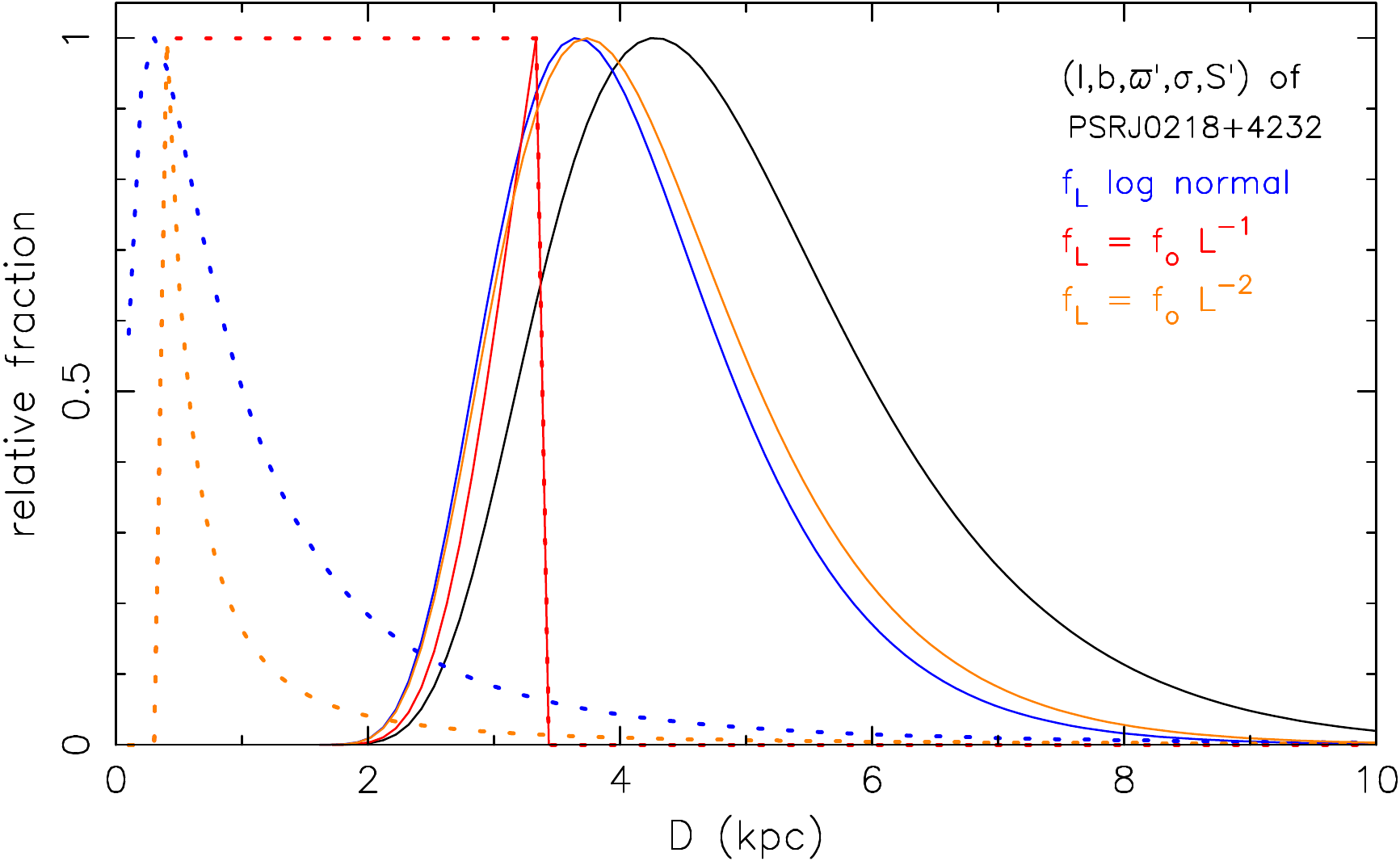}}

\caption{The distance probability function determined from parallax and
accurate flux, with known spatial and luminosity distributions. The black line
gives the values based on parallax only (reproducing the black line
in Fig.\ref{f:galacent}). For a lognormal luminosity function, the
dashed blue line shows the extra term $D^2f_L(L_oD^2S')$ and the blue 
solid line the overall distribution according to Eq.\ref{e:dparfluxb},
for values appropriate for \psrj.
The dashed and solid red and brown lines idem for  power-law luminosity functions
both with $L_\mathrm{min}=0.1$\,mjy\,kpc$^2$, and $L_\mathrm{max}=
10$\,mJy\,kpc$^{2}$ for $\alpha=-1$, and $L_\mathrm{max}=
100$\,mJy\,kpc$^{2}$ for $\alpha=-2$, respectively.
All curves are normalized to their maximum value.
% {\bf distflux.f90}
 \label{f:distflux}}
 \end{figure}

Suppose that we are interested in the probability distribution
for distances only. We note that,
for a finite measurement error, a range of luminosities contributes
to the probability of measuring $S'$. By integrating over the
luminosity, we find the joint probability of $D,\varpi',S'$
$$ P(D,\varpi',S') \propto g_D(\varpi'|D)f_D(D)\int g_S(S'|S[L,D])f_L(L)dL=$$
\begin{equation}
\phantom{M} g_D(\varpi'|D)f_D(D)\int
g_S(S'|S[L,D])f_L(L_oD^2S))\,L_oD^2dS
\label{e:parflux}\end{equation}
where we use the fact that $g_D(D)$ and $f_D(D)$ do not depend on $L$.

In many applications, the flux is measured much more accurately than
the parallax, in the sense that $\sigma_S/S\ll 1$.  In that
case, for a measurement error distribution $g_S$ according to
Eq.\ref{e:gauss-s}, only values of $S$ close to $S'$ contribute to the
integral over $S$ in Eq.\ref{e:parflux}, and $f_L$ is close to
constant in that small interval. Thus the factor
$f_L(L_oD^2S)L_oD^2=f_L(L_oD^2S')L_oD^2$ may be written outside
of the integral, and the remaining integral $\int g_S(S'|S)dS=1$.
With Bayes' theorem we then
obtain (cf.\ Eqs.\ref{e:bayes}, \ref{e:conditional}):
\begin{eqnarray}
p_D(D|\varpi',S')&=& C_D(0,D_\mathrm{max})
D^2\mathcal{F}(D)\times\nonumber\\
& &\exp\left[-{(1/D-\varpi')^2\over2\sigma^2}\right]
L_oD^2f_L(L_oD^2S')
\label{e:dparfluxb}\end{eqnarray}
Apart from normalization, the only difference with Eqs.\ref{e:normhomogen}
and \ref{e:jointgal}  is the extra term $D^2f_L(L_oD^2S')$.
For $f_L\propto L^{-1}$,
the extra term $D^2f_L$ is constant, and thus $p_D(D|\varpi',S')$
(Eq.\ref{e:dparfluxb}) is identical to $p_D(D|\varpi')$
, except for a
normalization constant, provided $L_\mathrm{min}<L<L_\mathrm{max}$.

In Fig.\ref{f:distflux} we apply eq. \ref{e:dparfluxb} to \psrj, for three
luminosity functions, where we set the uncertainty of
the measured flux to zero, for illustrative purpose .

For the power-law luminosity function Eq.\ref{e:powerlaw} with
$\alpha=-1$ we fix minimum and maximum luminosities at
0.1\,mJy\,kpc$^2$ and 10\,mJy\,kpc$^2$, respectively.  The accurate
flux then leads to minimum and maximum distances at: $D_\mathrm{min}
=\sqrt{L_\mathrm{min}/L_oS'}=0.33$\,kpc and $D_\mathrm{max} =
\sqrt{L_\mathrm{max}/L_oS'}=3.3$\,kpc.  For this luminosity function
 $p_D(D|\varpi',S') \propto p_D(D|\varpi')$
in the range $L_\mathrm{min}<L<L_\mathrm{max}$.

For a steeper power law with $\alpha=-2$, the extra term
$D^2f_L\propto D^{-2}$ enhances the probability of lower
distances and lowers the probability of large distances.
We show this for $L_\mathrm{max}=100$\,mJy\,kpc$^2$.

In Fig.\ref{f:distflux} we also show Eq.\ref{e:dparfluxb} for the
lognormal distribution, applied to \psrj, which apart from the
normalization is rather similar to the result for a power-law
luminosity distribution with $\alpha=-2$.

For all three luminosity functions, the lower range of allowed
distances is determined mainly by the parallax and its error.

\subsection{Distances: earlier derivations\label{s:lwrong}}

Verbiest et al.\, (2012) use the lognormal luminosity function
Eq.\ref{e:lognormal}.  Entering this in Eq.\ref{e:dparfluxb} we obtain
\begin{eqnarray}
p_D(D|\varpi',S') &=& C_D(0,D_\mathrm{max}) D^2\mathcal{F}(D)
\exp\left[-\,{(1/D-\varpi')^2\over2\sigma^2}\right]\times \nonumber \\ 
  && \exp\left[-\,{(\log\left[L_oD^2S'\right]-\mu_x)^2\over2{\sigma_x}^2}\right]{1\over
    S'}
\label{e:dlognormal}\end{eqnarray}
Comparing this with Eq.26 of Verbiest et al.\ (2012), we see that
the $1/S'$ term in Eq.\ref{e:dlognormal}  is there replaced with
$1/D$. This variant arises because their Eq.\,25 has $d\lambda/dD$ instead
of the correct $d\lambda/dS'$, analogous to the error leading to
Eq.\ref{e:wrong}. As a result, the probability of actual distance for
measured parallax and flux given by Verbiest et al.\ (2012, their
Eq.\,27), has a weighting factor $1/D^3$, absent in the correct
version of our Eq.\ref{e:dlognormal} (and omits the weighting factor
$1/S'$, which however drops out in the normalization).

\subsection{Luminosities\label{s:parlum}}

In the case where we are interested in luminosities only, we write
the joint probability of $L$, $\varpi'$
and $S'$, averaged over distances $D$, by integrating
Eq.\ref{e:jointall} over $D$. Substituting $D=\varpi^{-1}$,
and $D_\mathrm{max}=1/\varpi_\mathrm{min}$ this leads to
\begin{eqnarray}
P(L,\varpi',S') &\propto& f_L(L) \int_{\varpi_\mathrm{min}}^\infty
{\varpi^{-2}\mathcal{F}({1\over\varpi})\over\sqrt{2\pi}\,\sigma}
\exp\left[-{(\varpi-\varpi')^2\over2\sigma^2}\right] \times
\nonumber \\
& & {1\over\sqrt{2\pi}\sigma_S}
\exp\left[-{(L\varpi^2/L_o-S')^2\over2{\sigma_S}^2}\right]\,\varpi^{-2}d\varpi
\label{e:lfromds}\end{eqnarray}

\begin{figure}
\centerline{\includegraphics[angle=0,width=\columnwidth]{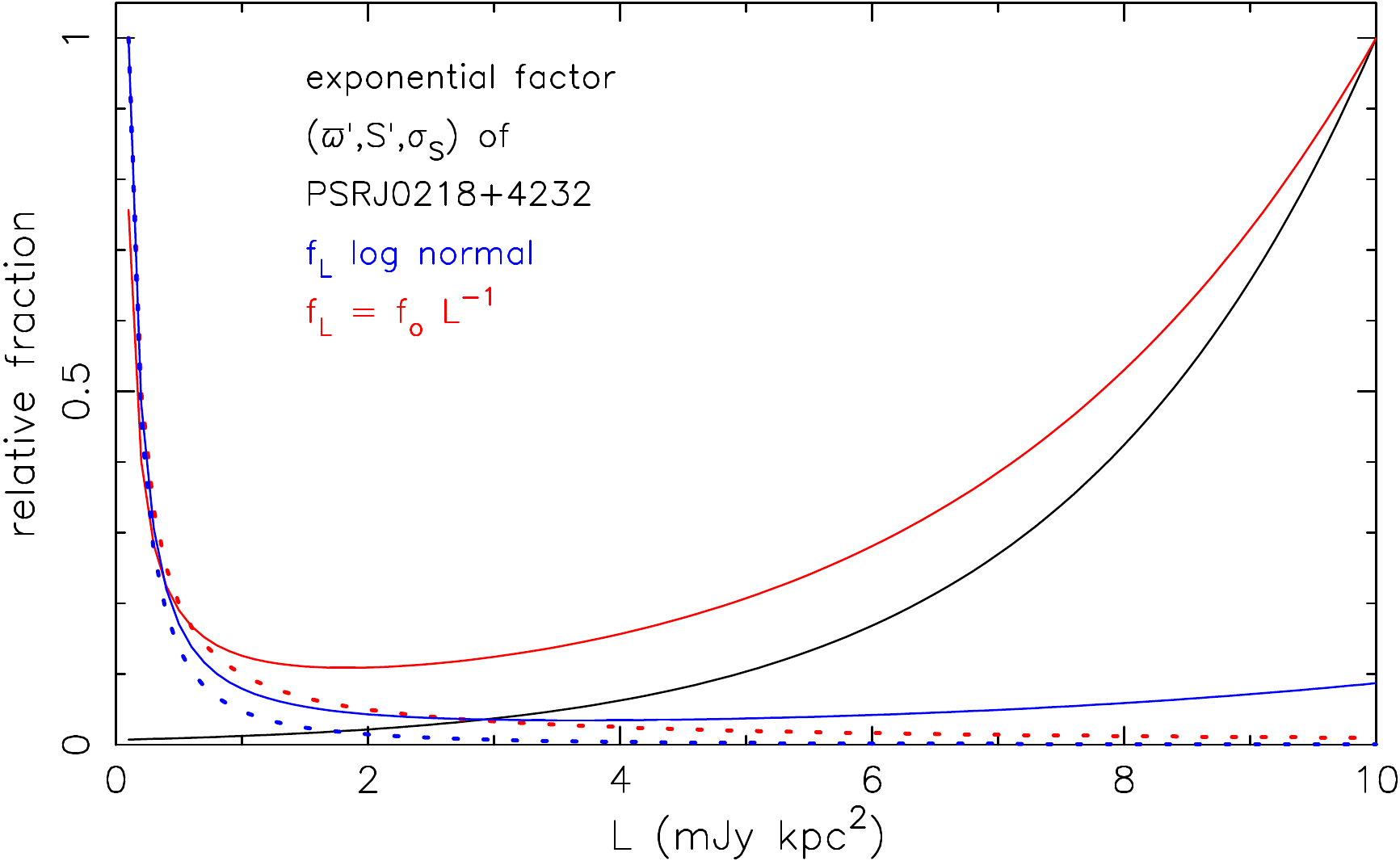}}

\caption{The luminosity probability function determined from accurate
parallax and uncertain flux (values for \psrj), with two 
assumed luminosity distributions. The black line
gives the exponential factor in Eq.\ref{e:lfromads}. The dashed red line
shows a power-law luminosity function, and the solid red line
the product of this with the exponential. The blue lines idem for
a lognormal luminosity function. Because of the accurate distance,
this luminosity probability function is valid for any
spatial distribution %{\bf lumacd.f90}
  \label{f:lumacd}}
 \end{figure}

\subsubsection{Luminosities with accurate distance}

We first consider the case where the distance is well known,
in the sense that $\sigma/\varpi\ll1$.
Only terms with $\varpi\simeq\varpi'$ contribute to the integral
over $\varpi$ in Eq.\ref{e:lfromds}, which may be rewritten as
\begin{eqnarray}
P(L,\varpi',S') &\propto& f_L(L) {\varpi'}^{-4}\mathcal{F}\left({1\over\varpi'}\right)
\exp\left[-{(L{\varpi'}^2/L_o-S')^2\over2{\sigma_S}^2}\right]\times
\nonumber \\
& &  \int_{\varpi_\mathrm{min}}^\infty
{1\over\sqrt{2\pi}\,\sigma}\exp\left[-{(\varpi-\varpi')^2\over2\sigma^2}\right] \times
d\varpi
\end{eqnarray}
The integral is a constant for a given
$\varpi_\mathrm{min}\equiv1/D_\mathrm{max}$,
and approaches unity when $\sigma/\varpi'$ approached zero, provided
$\varpi'>\varpi_\mathrm{min}$, i.e.\ provided that the nominal
distance $D'\equiv1/\varpi'$ satisfies $D'<D_\mathrm{max}$. We then have
\begin{eqnarray}
p_L(L|\varpi',S')& = & C_L(L_\mathrm{min},L_\mathrm{max})
{\varpi'}^{-4}\mathcal{F}({1\over\varpi'}) f_L(L) \nonumber \\
& & \times
\exp\left[-{[L-L'(\varpi',S')]^2\over2{\sigma_S(\varpi')}^2}\right]
\nonumber \\
\mathrm{where} & & L'(\varpi',S')={L_oS'\over{\varpi'}^2};\quad
\sigma_S'(\varpi') = {L_o\sigma_s\over{\varpi'}^2}
\label{e:lfromads}\end{eqnarray}
Because the integral over $L$ implicit in
$C_L(L_\mathrm{min},L_\mathrm{max})$ does not depend on $D$, the
factor $D^4\mathcal{F}={\varpi'}^{-4}\mathcal{F}({1/\varpi'})$ may
be dropped from this equation. Specifically, this implies that
$p_L(L|\varpi',S')$ does not depend on the spatial distribution.
Eq.\ref{e:lfromads} can be interpreted
directly, as follows. For an accurate distance $D=1/\varpi'$, the
number of sources scales with $D^2\mathcal{F}$. An extra factor $D^2$
is due to the conversion of a flux interval $\Delta S$ to a luminosity
interval $\Delta L$.  The probability of luminosity $L$ is given by
the probability of the corresponding flux $S=L/(L_oD^2)$, weighted
with the luminosity function $f_L$.  The weighting factor $f_L$ in
general may cause the most probable luminosity to differ from the
nominal luminosity $L'=L_oD^2S'$ -- analogous to the way in which the
weighting factor $f_D$ causes the most probable distance to differ
from the nominal distance $D'=1/\varpi'$ in Eqs.\ref{e:normhomogen}
and Eq.\ref{e:jointgal}.

The effect of the competition between the luminosity function $f_L$
and the exponential term in Eq.\ref{e:lfromads} can be quite dramatic,
as illustrated in Fig.\ref{f:lumacd}.
As an example we consider \psrj, assuming for illustrative
purpose that its parallax is exact. With 
$L_o=1$, its nominal luminosity is
$L'=L_oS'/{\varpi'}^2\simeq35$\,mJy\,kpc$^2$, and 
$\sigma_S'\simeq8$\,mJy\,kpc$^2$.
In the luminosity range considered, $0.1<L (\mathrm{mJy kpc}^2) <10$,
the exponential factor in Eq.\ref{e:lfromads} increases by a factor
130 between the low and the high luminosity limit.
The relatively flat luminosity function $f_L\propto L^{-1}$
decreases by a factor 100 in the same range. As a result, the 
overal luminosity probability peaks both at 0.1\,mJy\,kpc$^2$, and
-- less steeply -- at 10\,mJy\,kpc$^2$. 

The peak at the high luminosity limit is lowered for a luminosity
function that drops faster towards high luminosities, as illustrated
in Fig.\ref{f:lumacd} for the lognormal distribution
Eq.\ref{e:lognormal}.  For a power law $f_L\propto L^{-2}$ the peak at
10\,mJy\,kpc$^2$ disappears.  On the other hand, if the flux
measurement error is halved from its actual value to
$\sigma_S=0.1$\,mJy, both power-law distributions and the lognormal
distribution all combine with the exponential function to give a peak
only at 10\,mJy\,kpc$^2$ in the relative probability.

\subsubsection{Luminosities with accurate flux}

\begin{figure}
\centerline{\includegraphics[angle=0,width=\columnwidth]{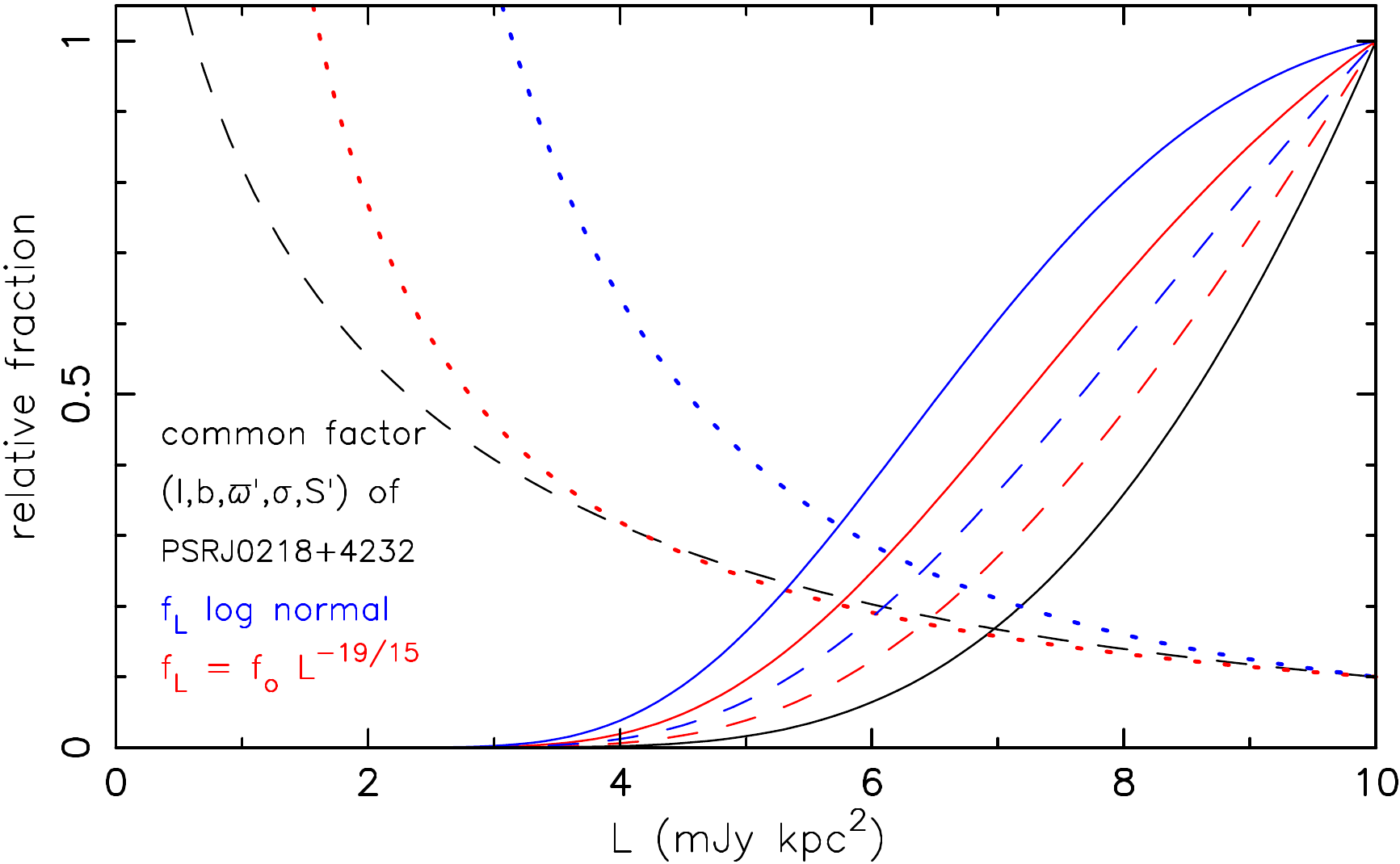}}

\caption{The luminosity probability function determined from uncertain
parallax and accurate flux (values for \psrj).  The black solid line shows
the exponential factor in Eq.\ref{e:lumacs}, multiplied with
$L^{1.5}$, the black dashed line shows $\mathcal{F}$. 
In blue, the dotted curve shows the 
lognormal luminosity distribution, the dashed and solid curves the
corresponding luminosity probability functions for homogeneous 
and galactocentric spatial distributions, respectively.
In red, idem for the power-law luminosity function.
All curves are normalized to a value 1 at 10\,kpc, except those
for the luminosity functions and for $\mathcal{F}$, normalized to a value of 0.1 at 10\,kpc
% {\bf lumacs.f90}
 \label{f:lumacs}}
 \end{figure}

To compute the integral in Eq.\ref{e:lfromds} in the limit
$\sigma_s/S'\ll1$, we first make the substitution
$\varpi^2=uL_o/L$, hence $2\varpi d\varpi=du\,L_o/L$.
Only terms with $u\simeq S'$ contribute to the integral,
hence
\begin{eqnarray}
P(L,\varpi',S') &\propto & f_L(L) x^{-5}\mathcal{F}\left({1\over
x}\right)\exp\left[-{(x-\varpi')^2\over2\sigma^2}\right] \times \nonumber\\
& & \int_{u_\mathrm{min}}^\infty\exp\left[ - {(u-S')^2\over2{\sigma_S}^2}\right]
du \,{L_o\over L} \nonumber\\
\mathrm{where} & & x = \sqrt{L_oS'\over L}; \quad \mathrm{and} \quad
u_\mathrm{min}={\varpi_\mathrm{min}}^2{L\over L_o} 
\end{eqnarray}
The integral depends on $L$, via $u_\mathrm{min}$. 
However, provided that $S'>u_\mathrm{min}$, i.e.\
$L<L_o{D_\mathrm{max}}^2S'$,
the integral approaches unity when $\sigma_s/S'$ approaches zero.
For $L>L_o{D_\mathrm{max}}^2S'$ the integral approaches zero in the
same limit.
Thus
\begin{equation}
  p_L(L|\varpi',S') =  \left\{ \begin{array}{l}
  C_L(L_\mathrm{min},L_\mathrm{max})f_L(L) \left({L\over
      L_oS'}\right)^{5/2}{L_o\over L}\mathcal{F}\left(\sqrt{L\over
      L_oS'}\,\right)\times \\ 
\phantom{m} \exp\left[-{(\sqrt{L_oS'/L}-\varpi')^2\over2\sigma^2}\right]\,
  \mathrm{for} \, L<L_o{D_\mathrm{max}}^2S' \\
\\
0 \quad  \mathrm{for} \quad L>L_o{D_\mathrm{max}}^2S'
\end{array}\right. 
\label{e:lumacs}
\end{equation}
Because the integral over $L$ implicit in
$C_L(L_\mathrm{min},L_\mathrm{max})$ does not depend on $L_o$ or $S'$,
the factor $L_o(L_oS')^{-5/2}$ may be omitted from Eq.\ref{e:lumacs}.

For the flux and parallax of \psrj, the exponential factor in
Eq.\ref{e:lumacs} increases with $L$ up to 10\,mJy\,kpc$^2$ (and beyond),
and this increase is amplified by the factor $L^{1.5}$. In contrast,
the luminosity functions increase towards the minimum luminosity
of 0.1\,mJy\,kpc$^2$. The combined effect of these two factors is
shown in Fig.\ref{f:lumacs}. The luminosity probability for the
galactocentric distribution observed in the direction of \psrj,
has a stronger contribution at luminosities below  10\,mJy\,kpc$^2$
than the homogeneous distribution. This is due to the rise
of $\mathcal{F}$ towards lower distances, hence lower luminosities.

\subsubsection{Can we do without the luminosity function?\label{s:alter}}

Since the luminosity is given by $L=L_oD^2S'$, one may wonder 
whether, in the case of accurate flux,  the probability of 
luminosity follows the probability of distance squared:
\begin{equation}
p_L(L|\varpi',S') \izzit K \,p_{D^2}(D^2|\varpi') 
\label{e:alter}\end{equation}
where $K$ is a proportionality constant.
The answer is no. 
This is most easily seen if we consider a standard candle,
where the luminosity function is unity for $L=L_S$ and
zero for all other luminosities $L\neq L_S$. An accurate flux then
implies that only one distance is possible, viz.\ the one
for which $L_oS'D^2=L_S$, whereas the right hand side of
Eq.\ref{e:alter} gives a non-zero value for a range of distances.

More generally,  at fixed flux $S'$ a different part of the 
luminosity function $f_L$ is sampled at different distances,
and thus the luminosity function is indispensable in the
determination of probabilities.

The invalidity of Eq.\ref{e:alter} implies that 
%a probability of the luminosity 
the probability density function of the luminosity
can be given only when the
luminosity function is known or assumed, or alternatively
when also the parallax is very accurate.

\section{The distance and gamma-ray luminosity of \psrj\label{s:gamma}}

\begin{figure}
\centerline{\includegraphics[angle=0,width=\columnwidth]{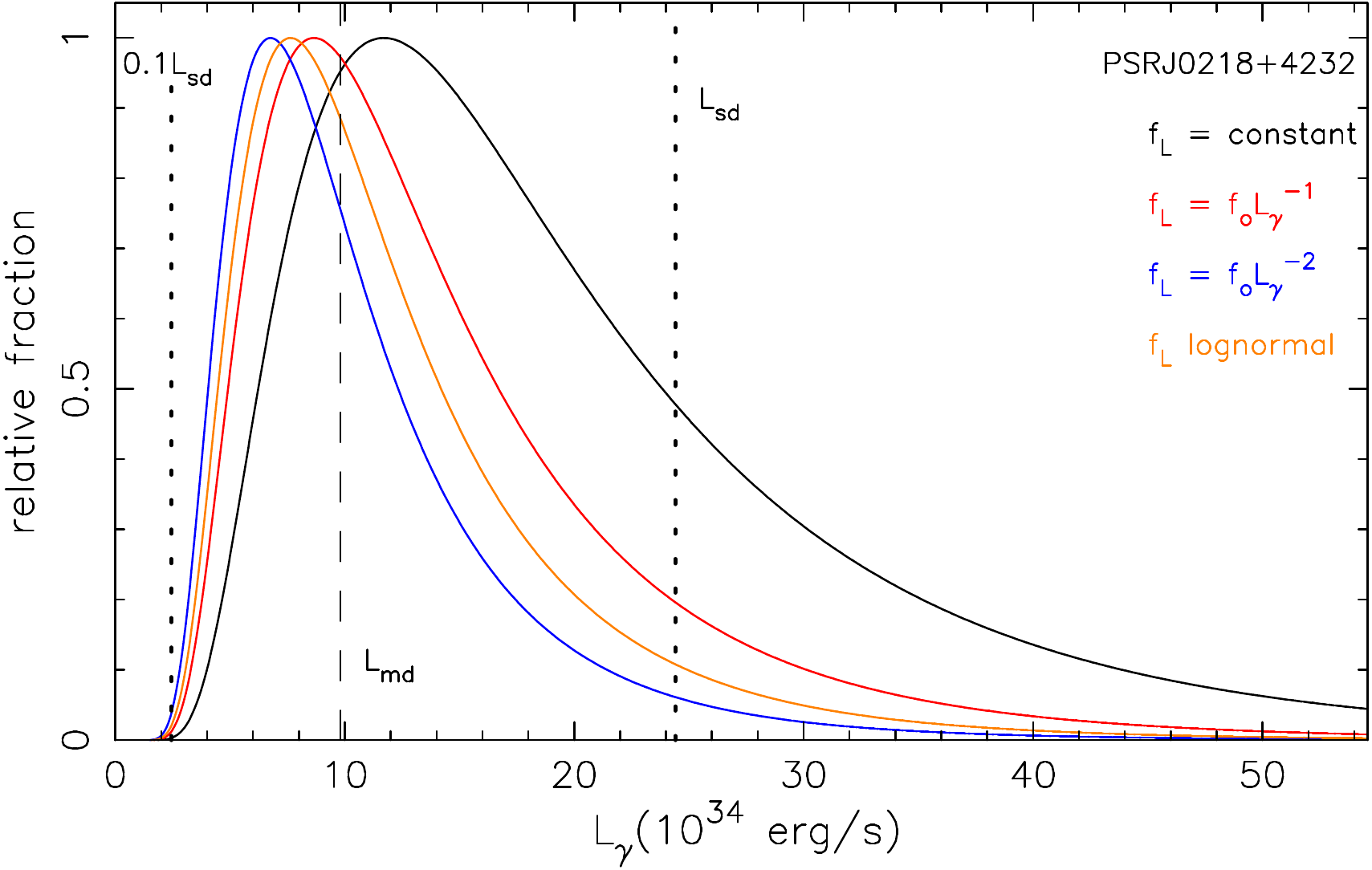}}

\caption{The gamma-ray luminosity probability function determined from the
parallax with error and accurate gamma-ray flux  of \psrj, for three different assumed
power-law and one lognormal gamma-ray luminosity functions.
The realistic distance prior (Eqs. \ref{e:spatial}, \ref{e:galacent}) is assumed. 
$L_\mathrm{sd}$ (Eq.~\ref{e:lsd}), and $0.1L_\mathrm{sd}$ are indicated with vertical dotted lines, 
$L_\mathrm{md}$  (Eq.~\ref{e:lgmd}) is indicated with a vertical dashed line. 
% {\bf lumgamma.f90}
 \label{f:gamma}}
 \end{figure}

\begin{figure}
\centerline{\includegraphics[angle=0,width=\columnwidth]{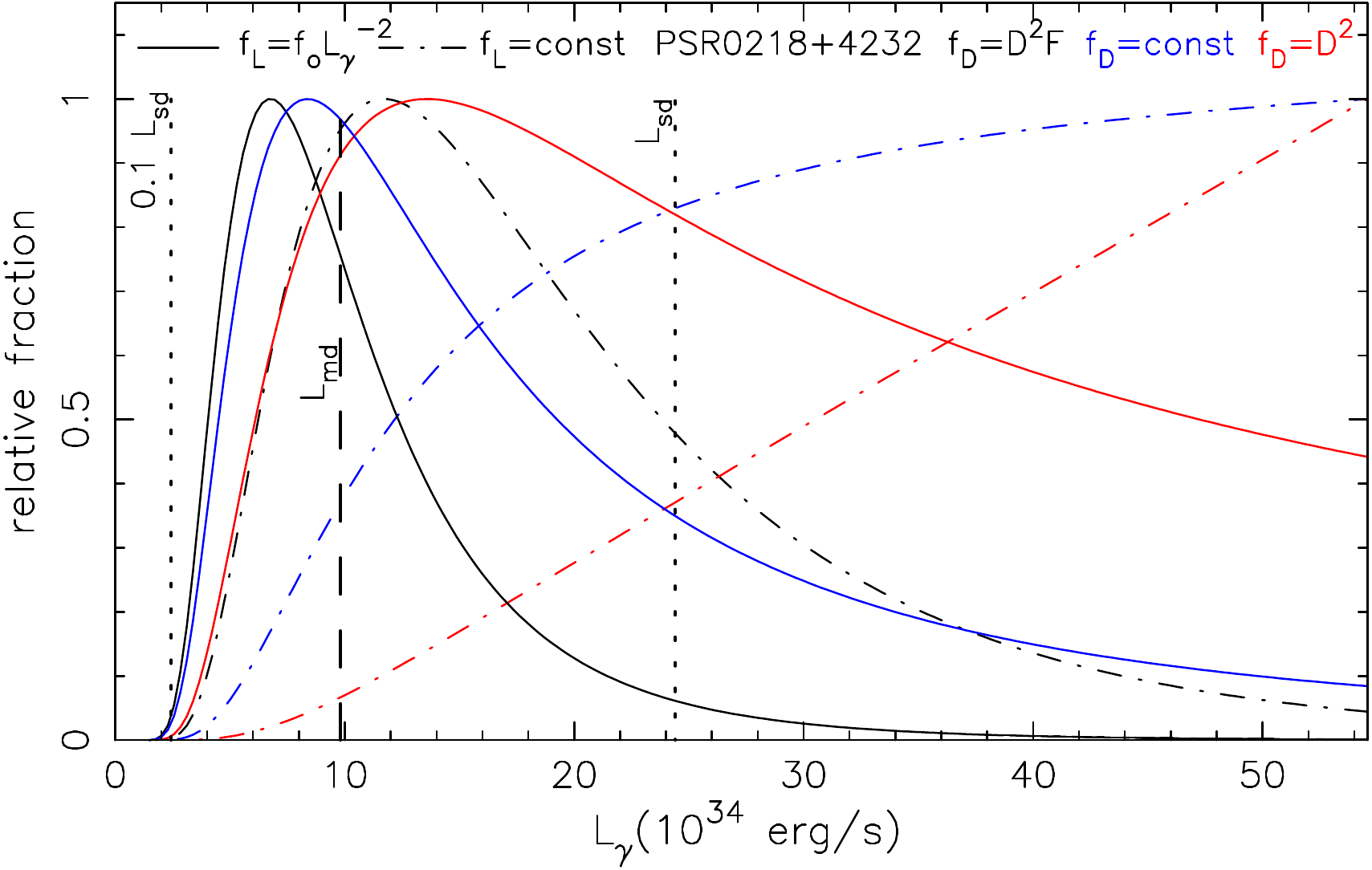}}

\caption{
The gamma-ray luminosity probability function determined from the
parallax with error and accurate gamma-ray flux  of \psrj, for three different distance
priors and two different luminosity priors.
$L_\mathrm{sd}$ (Eq.~\ref{e:lsd}), and $0.1L_\mathrm{sd}$ are indicated with vertical dotted lines, 
$L_\mathrm{md}$  (Eq.~\ref{e:lgmd}) is indicated with a vertical dashed line. 
%power-law gamma-ray luminosity functions. The spindown luminosity
%$L_\mathrm{sd}$ is indicated with a vertical dotted line.
% {\bf lumgamma.f90}
 \label{f:dgamma}}
 \end{figure}

An upper limit to the rotation-powered gamma-ray luminosity $L_\gamma$
is given by the spindown luminosity $L_\mathrm{sd}$
\begin{equation}
L_\mathrm{sd} \equiv {4\pi^2I \dot P\over P^3} \simeq 2.44\times10^{35} \mathrm{erg/s}
\label{e:lsd}\end{equation}
where the numerical value is for \psrj\ (see Table~\ref{t:psrj}), with
an assumed moment of inertia $I=10^{45}\mathrm{g\,cm}^2$ for the
neutron star.  It should be noted that the moment of inertia $I$ of
the neutron star \psrj\ is uncertain, as its mass and radius are
uncertain.

As reference values we use the nominal gamma-ray luminosity, 
at the nominal distance  $D_\mathrm{n}\equiv1/\varpi'=6.25$\,kpc:
\begin{equation}
L_\mathrm{n} \equiv 4\pi {1\over{\varpi'}^2}\,S_\gamma 
= 2.1\times10^{35}\mathrm{erg/s}=0.87L_\mathrm{sd}
\label{e:lgnom}\end{equation}
and the luminosity at the most probable distance according to 
Eq.{19}, $D_\mathrm{md}=4.28$\,kpc
\begin{equation}
L_\mathrm{md} \equiv 4\pi {D_\mathrm{md}}^2\,S_\gamma 
= 9.8\times10^{34}\mathrm{erg/s}=0.40L_\mathrm{sd}
\label{e:lgmd}\end{equation}
Note that the gamma-ray luminosity is defined for isotropic emission
(i.e.\ $S_o=4\pi$), which the gamma pulsations show to be false.

As noted in the previous section, the probability distribution
of luminosity for measured parallax with error and accurate
flux, can be given only when a luminosity function is known or
assumed. 

The effect of using different priors is illustrated in Figures~\ref{f:gamma} and \ref{f:dgamma}
and in Tables~\ref{t:dist} and \ref{t:lum}.

Table~\ref{t:dist} shows that use of a realistic distance prior, $f_D \propto D^2 \mathcal{F}$ with $\mathcal{F}$ given by 
Eq.~\ref{e:galacent},  reduces the
most probable distance to a value smaller than for $f_D$ uniform or spatially homogeneous,
also when $f_L$ is implemented. Application of realistic luminosity priors narrows the $95\%$ distance credibility
interval, in particular when an upper bound to $L_\gamma$ is set equal to $L_\mathrm{sd}$. 
In this case the upper limit of the credibility interval is close to the nominal distance of $6.25$~kpc.
%all priors give different maximum probabilities. 

Figure~\ref{f:gamma} shows the probability density functions of $L_\gamma$ for
the realistic distance prior $f_D \propto D^2\mathcal{F}$ with $\mathcal{F}$ given by
Eq.~\ref{e:galacent}. For each luminosity prior the probability that $L_\gamma < L_\mathrm{sd}$
is very small, $< 0.001$, and the most probable luminosity $L_\mathrm{mp}$ is well above 
$0.1 L_\mathrm{sd}$ and well below $L_\mathrm{sd}$.  
For steeper luminosiy functions the probability density function is pushed to lower luminosities,
as expected (see Table~\ref{t:lum}). 
%Use of realistic luminosity priors 

%Obviously, it differs from each of the most probable luminosities that
%is derived for each of the four luminosity priors shown, and
%more importantly is {\em not} the same as using a
%flat prior ($f_L=\mathrm{constant}$) in Eq.\ref{e:lumacs}.
%Also, a single value cannot represent a probability distribution.

The influence of the distance prior is much more significant.  
%In addition to the realistic prior $f_D\propto D^2\mathcal{F}$,
%we apply the uniform prior $f_D=\mathrm{constant}$
%and the spatially homogeneous prior $f_D\propto D^2$
%with no boundary. The results are shown in Figure~\ref{f:dgamma}. 
%The realistic distance prior leads to a probability for $L_\gamma$ which
%peaks at $L_\gamma < L_{sd}$. 
The unrealistic distance priors,
combined with the large uncertainty in the parallax, lead to 
unreallistically high $L_\gamma > L_{sd}$, especially for the uniform luminosity prior.

\begin{table}
\caption{The most probable distance $D_\mathrm{mp}$ and $95\%$ credibility interval for different distance and luminosity priors for \psrj.
}
\label{t:dist}
\begin{center}
\begin{tabular}{llccccc}
\hline
\multicolumn{2}{l}{Priors}            & $D_\mathrm{mp}$ & $D_l$ -- $D_u$ (a)  & $D_l$ -- $D_u$ (b)     \\
$f_D$  & $f_L$                        & (kpc)	        &   (kpc)  &  (kpc)             \\ 
\hline
const	           & --                & 6.25      & 3.75 -- 10.0     &  --    \\	    
$D^2$              & --                & 10.0          & 4.62 -- 10.0 &  --      \\
$ D^2 \mathcal{F}$ & --                & 4.28    &         2.65 --  7.82&   --  \\
$D^2 \mathcal{F}$  & lognorm           & 3.99  & 2.51 -- 7.15 & 2.71 --  6.38     \\
$D^2 \mathcal{F}$  & $\mathrm{const}$  & 5.05 & 3.08 -- 8.95   & 3.39 --  6.69     \\
$D^2 \mathcal{F}$  & $ L^{-1}$         & 4.28  & 2.65 -- 7.82 & 2.97 --  6.59     \\
$D^2 \mathcal{F}$  & $ L^{-2}$         & 3.74  & 2.39 -- 6.65 & 2.50 --  6.15     \\ 
\hline
\end{tabular}
\tablefoot{
All distance priors have  $D_\mathrm{max} = 10$ kpc; $\mathcal{F}$ refers to eq.~\ref{e:galacent}.
The luminosity priors are for gamma rays.  
 A dash means no priors applied: note that this is different from using a uniform prior $f_L = \mathrm{constant}$.
(a) No upper bound imposed on luminosity prior. (b) Maximum luminosity is $L_\mathrm{sd}$.
}
\end{center}
\end{table}

\begin{table}
\caption{The most probable gamma-ray luminosity $L_\mathrm{mp}$ and $95\%$ credibility interval for different distance and luminosity priors for \psrj.
}
\label{t:lum} 
\begin{center}
\begin{tabular}{l@{\hspace{0.06cm}}l@{\hspace{0.06cm}}cccc}
\hline
\multicolumn{2}{l}{Priors} & $L_\mathrm{mp}$ & $L_l$ -- $L_u$  & $\int_{L_\mathrm{sd}}^\infty p_L(L) dL$  \\
$f_D$ & $f_L$ & ($L_\mathrm{sd}$)  & ($L_\mathrm{sd}$)  \\
\hline
--                & --                &  0.87   & --               & -- \\ 
$D^2$             & --                &  2.24   & $0.70$ -- $2.24$ & $0.86$ \\
$\mathrm{const}$  &  $\mathrm{const}$ &  2.24   & $0.48$ -- $2.24$ & $0.72$ \\
$D^2$             & $L^{-2}$          &  0.56   & $0.24$ -- $2.14$ & $0.51$ \\
$D^2\mathcal{F}$  & $\mathrm{const}$  &  0.48   & $0.17$ -- $1.72$ & 0.26 \\
$D^2\mathcal{F}$  & $L^{-1}$          &  0.36   & $0.13$ -- $1.32$ & 0.12 \\
$D^2\mathcal{F}$  & $L^{-2}$          &  0.28   & $0.11$ -- $0.96$ & 0.04 \\
$D^2\mathcal{F}$  &  lognorm          &  0.31   & $0.12$ -- $1.11$ & 0.07 \\
\hline
\end{tabular}
\tablefoot{
A dash means no priors applied: note that this is different from using a uniform prior. 
All distance priors have upper boundary $D_\mathrm{max} = 10$ kpc: $\mathcal{F}$ refers to eq.~\ref{e:galacent}.
}
\end{center}
\end{table}

\section{Conclusions and discussion\label{s:discussion}}

A homogeneous spatial distribution is useful for pedagogical purposes
in explaining the importance of a prior in deriving a distance
probability distribution from a measured parallax.  For realistic
investigations, however, a homogeneous spatial distribution is rather
misleading. In particular, for a homogeneous spatial distribution, the
number of sources increases with distance, and a measured parallax
will more often correspond to a large distance which is measured too
low, than to a small distance measured too high.  In this case, a
parallax more often underestimates the actual distance, especially for
large measurement uncertainties (see Fig.\ref{f:homogenb}). In a
realistic galactic distribution, as observed from Earth, a parallax
tends to overestimate the distance, however, at distances where the
intrinsic source distribution decreases with distance
(Fig.\ref{f:galacent}). This is often the case, for example in
directions away from the galactic center and / or away from the
galactic plane.

Both analytically and via a Monte Carlo simulation, we show that
a prior for the spatial distribution must be used, also in the study
of a single object, for the determination of the distance probability density.
Similarly, when parallax and flux measurements with their
errors are combined to derive probability density distributions
for distances and luminosities, priors are necessary for both
spatial and luminosity distributions.
The nominal distance $D'=1/\varpi'$ and luminosity
$L'=L_oS/{\varpi'}^2$ may be very different from the
most probable values (see Fig.\ref{f:dandl}), unless both
measurement erors are small. 
This is the consequence of the predominance of low
luminosities in the luminosity functions that we use: 
for each flux the
higher probability of a low luminosity translates into a
higher probability of a lower distance -- in as far as
the parallax measurement allows this. In the case of \psrj,
for example, the most probable distance as derived fom the parallax
only is at 4.28\,kpc (Fig.\ref{f:distflux}). When parallax and radio flux
are both used, the most probable distance drops to 3.74\, kpc and
3.42\,kpc for power-law luminosity functions
with index $\alpha=-1$ and $\alpha=-2$, respectively; and
to 3.25\,kpc for the lognormal luminosity distribution
(Fig.\ref{f:dandl}). 
Clearly, the quality of the estimates of 
distance and luminosity is enhanced by the use of realistic prior 
distributions with respect to the nominal estimates based on 
measurement only. On the other hand, it is important to keep in 
mind that wrong priors may deteriorate the estimate. 

%\textbf{To illustrate the effect of varying priors, we list in
%Table~\ref{t:dist} the most probable distances and gamma-ray luminosities with
%their 95\%\ credibility intervals, for different combinations of
%distance and luminosity priors, as applied to \psrj.}
In particular the use of the spatial homogeneous prior is harmful in the case 
of an uncertain parallax: it shifts the value for most probable
distance or luminosity to the upper boundary on the prior (see second line
 in Tables 2 and 3). In contrast, the realistic
 distance prior gives an estimate for the gamma-ray luminosity inside the physically
 motivated region ($L_\mathrm{mp} < L_\mathrm{sd}$) even when no additional restrictions on the luminosity function are imposed.
An application of the lognormal luminosity prior gives an estimate for distance 
and gamma-ray luminosity which is in between two values obtained if we apply power law with $\alpha=-1$ and $\alpha=-2$.

It may be noted, in particular for the power-law luminosity function,
that the luminosity function may have a different form at different
luminosities (for an example,
see Eq.17 of Faucher-Gigu\`ere \&\ Kaspi 2006.) This is easily
implemented in the formalism described in the previous Sections.  More
complicated is the -- probably realistic -- case where the luminosity
function depends on the position in the Galaxy.  For millisecond
pulsars this is unlikely. Ordinary pulsars at large $z$, however, are
on average older than pulsars close to the galactic plane, and may
well have lower luminosities, if the pulsar luminosity depends on its
period and / or period derivative. For the study of such pulsars
an evolutionary model is indispensable in the determination
of their distances and luminosities.

In the study of a single object, the priors of spatial and luminosity
distributions must be known. In the study of a larger number of
objects, however, these distributions can and indeed should be derived
from prior observations. In general one may still wish to describe these
distributions with a number of parameters, e.g. $H$ and $h$ in
Eq.\ref{e:galacent}, $\alpha$ in Eq.\ref{e:powerlaw}, or $\mu_x$ and
$\sigma_x$ in Eq.\ref{e:lognormal}. 
For a sufficiently large number of pulsars, the evolutionary model
can also be tested.
At the moment, such studies are hampered by the lack of reliable
large ($>1$\,kpc) distances.

\begin{acknowledgements}
We thank Gijs Nelemans for discussions and suggestions. 
The research of AI is supported by a NOVA grant.
\end{acknowledgements}

\bibliographystyle{aa}
\bibliography{dist}

\end{document}